\documentstyle[psfig,conf_iap]{article}
\begin{document}

\heading{%
%Begin Heading
%
Limits on the Variability of Physical Constants
%
%End Heading
} 
\par\medskip\noindent
\author{%
%Begin Author names
M.J. Drinkwater$^1$, J.K. Webb$^1$, J.D. Barrow$^2$, V.V. Flambaum$^1$
%End Author names
}
\address{%
%First address
School of Physics, University of New South Wales, Sydney 2052, Australia
}
\address{%
% Second Address
Astronomy Centre, University of Sussex, Brighton, BN1 9QH, UK
}

\begin{abstract}
%Begin Abstract
We have compared the frequency of high-redshift Hydrogen HI 21\,cm
absorption with that of associated molecular absorption
in two quasars to place new (1 sigma) upper limits on any
variation in $y \equiv g_p\alpha^2$ (where $\alpha$ is the fine
structure constant and $g_p$ is the proton g-factor)
of $\Delta y / y < 5 \times 10^{-6}$ at
redshifts $z=0.25$ and $z=0.68$.
%End Abstract
\end{abstract}

\section{Introduction}

Recent measurements of molecular absorption in some radio sources
corresponding to known HI 21\,cm absorption systems give us ideal
laboratories in which we can search for any possible temporal and
spatial variation in the fundamental constants of Nature (a summary is
given in \cite{V95}).  The rotational transition frequencies of
diatomic molecules such as CO are proportional to ${\hbar /( Ma^2)}$
where $M$ is the reduced mass and $a={\hbar^2 /( m_e e^2)}$ is the
Bohr radius. The 21\,cm hyperfine transition in hydrogen has a
frequency $\sim {\mu_p \mu_B /( \hbar a^3)}$, where $\mu_p = g_p {e
\hbar / (4m_p c)}$, $g_p$ is the proton g-factor and $\mu_B = {e
\hbar / (2m_e c)}$. Consequently (assuming $m_p/M$ is constant) the
ratio of the hyperfine frequency to the molecular rotational frequency
$\sim g_p \alpha^2$ where $\alpha = {e^2 / ( \hbar c)}$ is the fine
structure constant. Any variation in $y\equiv g_p \alpha^2$ would
therefore be observed as a difference in the apparent redshifts:
${\Delta z / (1+z)} \approx {\Delta y / y}$. Two sources
have common extragalactic molecular and HI absorption detections
published: 0218$+$357 and 1413$+$135 (see Table~1) which
we use in this paper to put constraints on any such variations.

\section{New HI Analysis}

One possible problem with this approach is that the HI and molecular
absorptions might result from different clouds along the respective
lines of sight. We investigated this by looking at a sample of
Galactic mm-wave continuum sources \cite{LL96} with both HCO$^+$
and HI \cite{D83} absorption detected. We deconvolved the complex HI
absorption into single systems and then identified the closest
matching HI line to each HCO$^+$ absorption: the distribution of
velocity differences is centred close to zero (mean $\Delta
v=0.4{\rm\,km\,s}^{-1}$) and very narrow with a
Gaussian dispersion of only 1.2\,km\,s$^{-1}$.  This is equal
to the spectral resolution of the HI data (0.6--1.3\thinspace
km\,s$^{-1}$) and shows the absorption is from the same
clouds to within our velocity resolution.
\raisebox{-2cm}[0cm][0cm]{\makebox[0cm][c]
{\parbox{14cm}{\em Paper presented at the 13$^{th}$ IAP
Colloquium: Structure and Evolution of the IGM from QSO Absorption
Line Systems, held in Paris, 1997 July 1-5, eds. P. Petitjean,
S. Charlot
}}}

We used the same approach to decompose the high-redshift HI
absorption systems listed in Table~1, fitting a number of components
to each system \cite{DWB}. The fitting routine gave uncertainties less than
1\,km\,s$^{-1}$ for the strong lines; these were added in
quadrature with the velocity scale uncertainty of 0.3\,km\thinspace
s$^{-1}$ to obtain the errors.  As with the low-redshift Galactic
absorption we picked the the HI absorption closest in velocity to
each molecular absorption and list these new redshifts in Table~1.

{\small
% Begin Table
\begin{center}
\begin{tabular}{rlrrc}
\multicolumn{5}{l}{{\bf Table 1. Comparison of Molecular and Atomic Absorption Data} } \\
\hline
%\\
\multicolumn{1}{c}{Name}      &\multicolumn{1}{c}{Molecular z} & \multicolumn{1}{c}{Atomic z} &\multicolumn{1}{c}{Atomic z (new)}&{$\sigma_{\Delta z / 1+z}$}\rule{0cm}{0.45cm}\rule[-0.25cm]{0cm}{0.3cm}  \\
\hline
%\\
		   
\rule{0cm}{0.45cm}
0218$+$357&$0.684680\pm$.000006\cite{WC95}& 0.68466\cite{C93}
                                           & $0.684684\pm$.000006 &5$\times10^{-6}$ \\
\rule[-0.25cm]{0cm}{0.3cm}
1413$+$135&$0.246710\pm$.000005\cite{WC94}& 0.24671\cite{C92}
                                           & $0.246710\pm$.000004 &5$\times10^{-6}$ \\
%1504$+$377&$0.673350\pm$.000005\cite{WC96}& 0.67324\cite{C97}
%                                           & $0.673236\pm$.000003 & --\\
%\rule[-0.25cm]{0cm}{0.3cm}
%1504$+$377&$0.673350\pm$.000005\cite{WC96}& 0.67343\cite{C97}
%                                           & $0.673423\pm$.000005 & --\\
%\\
\hline
\end{tabular}
\end{center}
%End Table
}

Molecular measurements are much more precise than the HI data
(velocity uncertainties of 0.1\,km\,s$^{-1}$) but the published values
were not quoted to this precision, so we estimated the molecular
redshifts and uncertainties from the publications as listed in Table~1
(we excluded a third source 1504$+$377 because of evidence of an
offset in its molecular velocity scale)\cite{DWB}. In both sources the
redshifts agree to within our errors so we combined the uncertainties
in quadrature to give 1 sigma upper limits on the redshift differences
in Table~1. The corresponding limits on any change in $y=g_p \alpha^2$
are ${\Delta y / y} < 5\times10^{-6}$ at $z=0.25$ and $z=0.68$. These
are significantly lower than the previous best limit of
$1\times10^{-4}$\cite{V96} (it was quoted as a limit on nuclueon mass,
but it actually refers to $g_p \alpha^2$). As there are no theoretical
grounds to expect that the changes in $g_p$ and $\alpha^2$ are
inversely proportional, we obtain independent rate-of-change limits of
$|\dot{g_p}/g_p| < 2\times10^{-15}{\rm\, y}^{-1}$ and
$|\dot{\alpha}/\alpha| < 1\times10^{-15}{\rm\, y}^{-1}$ at $z=0.25$
and $|\dot{g_p}/g_p| < 1 \times10^{-15}{\rm\, y}^{-1}$ and
$|\dot{\alpha}/\alpha| < 5\times10^{-16}{\rm\, y}^{-1}$ at $z=0.68$
(for $H_0=75$\,km\,s$^{-1}$\,Mpc$^{-1}$ and $q_0=0$).  These are much
lower than the previous  1 sigma limit of $|\dot{\alpha}/\alpha| <
8\times10^{-15}{\rm\, y}^{-1}$ at $z\approx 3$ \cite{VPI96}.

% Begin acknowledgements
\acknowledgements{We thank Chris Carilli, Harvey Liszt, and John Dickey for
many helpful discussions and for allowing us to use their data.}
% End acknowledgements

%References should be refered as : \cite{LH}, \cite{MMM}, and \cite{Kea}. 

%
\vfill
\end{document}